\documentclass[aps]{revtex4}
\usepackage{graphicx}
\begin{document}
\def\slash{\not\!}
\def\sslash{\not\!\!}
\def\vecalpha{{\mbox{\boldmath{$\alpha$}}}}
\def\vecnabla{{\mbox{\boldmath{$\nabla$}}}}
\def\vectau{{\vec\tau}}
\def\vecp{{\bf P}}
\def\vecr{{\bf r}}
\def\barpsi{\bar\psi}
\def\d{{\rm d}}
\title{Short range correlations in relativistic nuclear matter models}
\author{P.K. Panda, Jo\~ao da Provid\^encia and Constan\c{c}a Provid\^encia}
\affiliation{Departamento de F\'isica, Universidade de Coimbra, 3000
Coimbra, Portugal}
\begin{abstract}
Short range correlations are introduced using unitary correlation
method in a relativistic approach to the equation of state of the
infinite nuclear matter in the framework of the Hartree-Fock
approximation. It is shown that the correlations give rise to an
extra node in  the ground-state wave-function in the nucleons,
contrary to what happens in non-relativistic calculations with a
hard core. The effect of the correlations in the ground state
properties of the nuclear matter and neutron matter is studied. The
nucleon effective mass and equation of state (EOS) are very
sensitive to short range correlations. In particular, if the pion
contact term is neglected a softening of the EOS is predicted.
Correlations have also an important effect on the neutron matter EOS
which presents no binding but only a very shallow minimum contrary
to the Walecka model.

\end{abstract}
\date{\today} \maketitle
We discuss the role of short range correlations in a relativistic
approach to the description of nuclear matter. Several procedures
may be used to introduce short range correlations into the model
wave function. All have something in common with the so called $e^S$
method or coupled-cluster expansion, introduced by Coester and
K\"ummel \cite{coesterkuemmel} and most elegantly and extensively
developed and applied to diverse quantal systems by R. Bishop
\cite{bishop}.

 In this short note, we consider the unitary operator method as
 proposed by Villars \cite {villars}, which automatically guarantees
 that the correlated state is normalized. The general idea of
 introducing short range correlations in systems with short range
 interactions exists for a long time \cite{jp,feldmeier} but has not
 been pursued for the relativistic case.

 Non-relativistic calculations based on realistic NN potentials
 predict equilibrium points which do not reproduce simultaneously the
 binding energy and saturation density. Either the saturation density
 is reproduced but the binding energy is too small, or the binding
 energy is reproduced at too high a density \cite{Coester}. In order
 to solve this problem, the existence of a repulsive potential or
 density-dependent repulsive mechanism \cite{3-body} is usually
 assumed. Due to Lorentz covariance and self-consistency,
 relativistic mean field theories \cite{walecka}  include
 automatically contributions which are equivalent to $n$-body
 repulsive potentials in non-relativistic approaches.  The
 relativistic quenching of the scalar field provides a mechanism for
 saturation, though, by itself it may lead to too small an effective
 mass and too large incompressibility of nuclear matter, a situation
 which is encountered in the Walecka model \cite{walecka}.

 In the non-relativistic case, we find a wound in the relative wave
 function if the vector interaction is stronger than the scalar
 interaction. However, this may not be the case if the interaction is
 strong enough for short distances. Then, the effective potential can
 actually turn attractive at short distances and the wave function
 may well have a node, as proposed by V. Neudatchin \cite{neudatchin}
 and advocated by S. Moszkowski \cite{moszkowski}, on the basis of
 the the quark structure of the nucleon. The situation with short
 range correlations may be more subtle than might be thought from a
 simple non-relativistic model. We show that a short range node in
 the relative wave function  may
 be encountered in relativistic models, although it remains to be
 seen to what extent the relativistic description simulates the quark
structure.

In non-relativistic models the saturation arises from the interplay
between a long range attraction and a short range repulsion, so
strong that it is indispensable to take short range correlations
into account. In relativistic mean field models, the parameters are
phenomenologically fitted to the saturation properties of nuclear
matter.  Although in this approach short range correlation effects
may be accounted for, to some extent, by the model parameters, it is
our aim to study explicitly the consequences  of actual short range
correlations. In a previous publication \cite{corr} we have
discussed the effect of the correlations in the ground state
properties of  nuclear matter  in the framework of the
Hartree-Fock approximation using an effective Hamiltonian derived
from the $\sigma-\omega$ Walecka model.  We have shown, for
interactions mediated only by sigma and omega mesons, that the
equation of state (EOS) becomes considerably softer when
correlations are taken into account, provided the correlation
function is treated variationally, always paying careful attention
to the constraint imposed by the ``healing distance" requirement. In
the present note we will work within the same approach and will include
also the exchange of pions and $\rho$-mesons.
Preliminary results of the present work have been presented in
\cite{proc}

We start by considering the effective Hamiltonian \cite{corr}
\begin{equation}
 H=\int \psi^\dagger_\alpha (\vec x)(-i\vec\alpha\cdot\vec\nabla
 +\beta M)_{\alpha\beta}\psi_\beta(\vec x)~d\vec x
 +\frac{1}{2}\int\psi_\alpha^\dagger (\vec x)\psi_\gamma^\dagger (\vec y)
 V_{\alpha\beta,\gamma\delta}(|\vec x -\vec y|)
 \psi_\delta(\vec y)\psi_\beta(\vec x)~d\vec x~d\vec y,\label{hamiltonian}
 \end{equation}
where  the exchange of  $\sigma$, $\omega$, $\rho$  and $\pi$ mesons
is taken into account, so that
\begin{equation}
V_{\alpha\beta,\gamma\delta}(r)= \sum_{i=\sigma, \omega,\rho,\pi}
V_{\alpha\beta,\gamma\delta}^i(r)
\end{equation}
with
$$
V_{\alpha\beta,\gamma\delta}^\sigma(r)=-\frac{g_\sigma^2}{4
\pi}(\beta)_{\alpha\beta}(\beta)_{\gamma\delta}\frac{e^{-m_\sigma
r}}{r},
$$
$$
V_{\alpha\beta,\gamma\delta}^\omega(r)=\frac{g_\omega^2}{4
\pi}\left(\delta_{\alpha\beta}\delta_{\gamma\delta}
-\vec\alpha_{\alpha\beta}\cdot\vec\alpha_{\gamma\delta}\right)\frac{e^{-m_\omega
r}}{r},
$$
$$
V_{\alpha\beta,\gamma\delta}^\rho(r)=\frac{g_\rho^2}{4
 \pi}\left(\delta_{\alpha\beta}\delta_{\gamma\delta}
-\vec\alpha_{\alpha\beta}\cdot\vec\alpha_{\gamma\delta}\right)
\vec\tau_1\cdot\vec\tau_2\frac{e^{-m_\rho r}}{r}.
$$
An interaction of the form \cite{bouyssy}
\begin{equation}
V_{\alpha\beta,\gamma\delta}^\pi(\vec r)
=\frac{1}{3}\left[\frac{f_\pi}{m_\pi}\right]^2
(\Sigma_i)_{\alpha\beta}(\Sigma_i)_{\gamma\delta}\vec\tau_1\cdot\vec\tau_2
\left[\frac{4\pi}{m_\pi^3}\delta(\vec r)-\frac{e^{-m_\pi r}}{m_\pi
r}\right], \label{pi}
 \end{equation}
 where $\Sigma_i=\alpha_i\gamma_5$, describes the exchange of $\pi$
 mesons. The first term in the above is the repulsive contact
 interaction and the second term is an attractive Yukawa potential.
 This can be rewritten in momentum space as
 \begin{equation}
 V_{\alpha\beta\gamma\delta}^\pi(\vec q)=
 \frac{1}{3}\left[\frac{f_\pi}{m_\pi}\right]^2
 (\Sigma_i)_{\alpha\beta}(\Sigma_i)_{\gamma\delta}~
 \vec\tau_1\cdot\vec\tau_2 \left[ \frac{q^2}{q^2+m_\pi^2}\right].
 \end{equation}

In equation (\ref{hamiltonian}), $\psi$ is the nucleon field
interacting through the scalar and vector potentials and
$\vec\alpha,\,\beta$ are the Dirac-matrices. The equal time
quantization condition for the
 nucleons reads,
$ [\psi _{\alpha}(\vec x,t),\psi _\beta (\vec y,t)^{\dagger}]_{+}
 =\delta _{\alpha \beta}\delta(\vec x -\vec y),
$
 where the indices $\alpha$ and $\beta$ refer to the spin. The field
 expansion for the field $\psi$ at time t=0 reads \cite{mishra}
 \begin{equation}
 \psi(\vec x)=\frac {1}{\sqrt{V}}\sum_{r,k} \left[U_r(\vec k)c_{r,\vec k}
 +V_r(-\vec k)\tilde c_{r,-\vec k}^\dagger\right] e^{i\vec k\cdot \vec x} ,
 \end{equation}
 where $U_r$ and $V_r$ are
 \begin{equation}
 U_r(\vec k)=\left( \begin{array}{c}\cos\frac{\chi(\vec k)}{2}
 \\ \vec \sigma \cdot\hat k\sin\frac{\chi(\vec k)}{2}
 \end{array}\right)u_r~ ;~~ V_r(-\vec k)=\left(
 \begin{array}{c}-\vec \sigma \cdot\hat k\sin\frac{\chi(\vec k)}{2}
 \\ \cos\frac{\chi(\vec k)}{2}\end{array}\right)v_r~.
 \end{equation}
 For free spinor fields, we have $\cos\chi(\vec k)=M/\epsilon(\vec
 k)$, $\sin\chi(\vec k)=|\vec k|/\epsilon(\vec k)$ with
 $\epsilon(\vec k)=\sqrt{\vec k^2 + M^2}$. However, we will deal with
 interacting fields so that we take the ansatz $\cos\chi(\vec
 k)=M^*(\vec k)/\epsilon^*(\vec k)$, $\sin\chi(\vec k)=|\vec
 k^*|/\epsilon^*(\vec k)$, with $\epsilon^*(\vec k)=\sqrt{\vec
 {k^*}^2 + {M^*}^2(\vec k)}$, where $\vec k^*$ and $M^*(\vec k)$ are
 the effective momentum and the effective mass, respectively,
 determined self-consistently by the Hartree-Fock (HF) prescription.
 The vacuum $\mid 0\rangle$ is defined through $c_{r,\vec k}\mid
 0\rangle=\tilde c_{r,\vec k}^\dagger\mid 0 \rangle=0$;
 one-particle states are written $|\vec k,r\rangle=c_{r,\vec
 k}^\dagger\mid 0\rangle$; two-particle and three-particle
 uncorrelated states are written, respectively as $|\vec k,r;\vec
 k',r'\rangle= c_{r,\vec k}^\dagger ~c_{r',\vec k'}^\dagger\mid
 0\rangle$, and $|\vec k,r;\vec k',r'; \vec k'',r''\rangle=
 c_{r,\vec k}^\dagger ~c_{r',\vec k'}^\dagger ~c_{r'',\vec
 k''}^\dagger \mid 0\rangle,$ and so on.

We now introduce the short range correlations through the unitary
 operator method. The correlated wave function \cite{jp1} is
 $|\Psi\rangle=e^{i\Omega}|\Phi\rangle$ where $|\Phi\rangle$ is a
 Slater determinant and $\Omega$ is, in general, a $n$-body Hermitian
 operator, splitting into a 2-body part, a 3-body part, etc.. The
 expectation value of $H$ is
 \begin{equation}
 E=\frac{\langle \Psi| H|\Psi\rangle}{\langle \Psi |\Psi\rangle}=
 \frac{\langle \Phi| e^{-i\Omega}~H~e^{i\Omega}|\Phi\rangle}{\langle \Phi|
 \Phi\rangle}.
 \end{equation}

 In the present calculation, we only take into account two-body
 correlations.
 Let us denote the  two-body  correlated wave function by
$ |\overline{\vec
 k,r;\vec k',r'}\rangle=e^{i\Omega}|{\vec k,r;\vec k',r'}\rangle\approx
 f_{12}|{\vec k,r;\vec k',r'}\rangle
$ where $f_{12}$ is the short range correlation factor, the
so-called Jastrow factor \cite{jastrow}. For simplicity, we consider
$f_{12}=f(\vec r_{12})$, $\vec r_{12}=\vec r_1-\vec r_2$, and
$f(r)=1-(\alpha +\beta r)~e^{-\gamma r}$ where $\alpha$, $\beta$ and
$\gamma$ are parameters. The choice of a real function for the
unitary operator has to be supplemented by a normalization condition
(see eq. (\ref{c0})) which assures unitarity to leading cluster
order.

 The important effect of the short range
 correlations is the expression for the correlated ground-state
 energy. Here, in the leading order of the cluster expansion, the
 interaction matrix element $\langle{\vec k,r;\vec
 k',r'}|V_{12}|{\vec k,r;\vec k',r'}\rangle$ of the HF expression is
 replaced by $\langle\overline{\vec k,r;\vec
 k',r'}|V_{12}+t_1+t_2|\overline{\vec k,r;\vec
 k',r'}\rangle-\langle{\vec k,r;\vec k',r'}|t_1+t_2|{\vec k,r;\vec
 k',r'}\rangle$, where $t_i$ is the kinetic energy operator of
 particle $i$. As argued by Moszkowski \cite{mosz} and Bethe
 \cite{bethe}, it is expected that the true ground-state wave
 function of the nucleus, containing correlations, coincides with the
 independent particle, or HF wave function, for inter particle
 distances $r\geq r_{h}$, where $r_{h}\approx 1$ fm is the so-called
 ``healing distance". This behavior is a consequence of the
 restrictions imposed by the Pauli Principle. A natural consequence
 of having the correlations introduced by a unitary operator is a
 normalization constraint on $f(r)$,
 \begin{equation}
 \int~(f^2(r)-1)~d^3r=0.\label{c0}
 \end{equation}

 The correlated ground state energy of symmetric nuclear matter reads
 \begin{eqnarray}
 {\cal E}&=&\frac{\nu}{\pi^2}\int_0^{k_F} k^2 ~dk~
 \left[|k|\sin\chi(k)~+~M\cos\chi(k)\right]
 ~+~\frac{\tilde F_\sigma(0)}{2}\rho_s^2 +\frac{\tilde
 F_{\omega}(0)}{2}\rho_B^2\nonumber\\
&-&\frac{4}{(2\pi)^4} \int_0^{k_f} k^2~ dk~ {k'}^2~ d
 k'~ \left\{\Big[ |k| \sin\chi(k)+ 2~M\cos\chi(k)\Big] I(k,k')
 + |k|~\sin\chi(k')~J(k,k') \right\}\nonumber\\
 &+&\frac{1}{(2\pi)^4}\int_0^{k_f} k~ dk~ k'~ dk'
 \left[\sum_{i=\sigma,\omega,\rho,\pi} A_i (k,k')
 +\cos\chi(k)\cos\chi(k') \sum_{i=\sigma,\omega,\rho,\pi}B_i
 (k,k')\right.\nonumber\\
 &+&\left. \sin\chi(k)\sin\chi(k')
 \sum_{i=\sigma,\omega,\rho,\pi} C_i(k,k')\right]\nonumber\\
 \end{eqnarray}
where $A_i$, $B_i$, $C_i$ , $I$ and $J$ are exchange integrals
defined in the appendix.  In the above equation, the first term
comes from the kinetic contribution, the second and third
terms come respectively from the $\sigma$ and  $\omega$ direct
contributions to the correlated potential energy, the other terms
arise from the exchange correlation contribution to the kinetic
energy, and  from the meson exchange contributions to the correlated
potential energy. The direct term of the correlation contribution to
the kinetic energy vanishes  due to (\ref{c0}), and $\rho_B$ and
$\rho_s$ are, respectively, the baryon and the scalar densities.

The couplings $g_\sigma$, $g_\omega$,  $g_\rho$, $g_\pi$, the meson
masses, $m_i,\, i=\sigma,\, \omega,\, \rho,\, \pi$ and the three
parameters specifying the short range correlation function,
$\alpha$, $\beta$ and $\gamma$ have to be fixed. The couplings
$g_\sigma$ and  $g_\omega$ are chosen so as to reproduce the ground
state properties of nuclear matter. For the $\rho$ and $\pi$-meson
couplings we take the usual values
  $g_\rho^2/4\pi=0.55$ and $f_\pi^2/4\pi=0.08$.
We choose $m_\sigma=550$ MeV,
 $m_\omega=783$ MeV$, m_\rho=770$ MeV and
 $m_\pi=138$ MeV. The normalization condition (\ref{c0})
 determines $\beta$. We fix $\alpha$ by minimizing the energy. The
 parameter $\gamma$ is such that it reproduces a reasonable healing
 distance, assuming that this quantity decreases as $k_F$ increases.
 Therefore, we assume that $\gamma$ depends on $k_F$ according to
 $\gamma=a_1+ a_2\, k_F/k_{F0}$, where the parameters $a_1$ and $a_2$
 are conveniently chosen.

 In tables \ref{tab} and \ref{tab1}, we have tabulated the parameters
 used in our calculation together with the relative effective mass
 $M^*/M$, the kinetic energy ${\cal T}/\rho_B-M$, the direct and
 exchange parts of the potential energy (${\cal V}_d/\rho_B$ and
 ${\cal V}_e/\rho_B$ respectively) with correlation, and the
 correlation contribution to the kinetic energy ${\cal T}^C/\rho_B$,
 all calculated at the saturation point. Notice that a HF calculation
 produces an EOS which is stiffer than the one obtained at the
 Hartree level.  However,  correlations reduce the effective mass and
 soften the EOS.  In fact, the contribution of direct and exchange
 correlation terms are of the same order of magnitude of the other
 terms in the energy per particle. Moreover, the values of the
 couplings $g_\sigma$ and $g_\omega$ which reproduce the saturation
 density and binding energy strongly depend on the correlations,
being considerably reduced by their presence, which is quite a
remarkable fact. Hence, short-range correlations cannot be
disregarded.

The correlation function $f(r)$ is plotted in figure 1 as a function
of the relative distance for two different situations:  in one
calculation the contact term coming from the pion contribution was
included and  the  other  curve was obtained excluding it. It is
clear from both curves that the correlations give rise to an extra
node in the dependence of the ground-state wave-function on the
relative coordinate, contrary to what generally happens in
non-relativistic calculations with a hard core, when the wave
function acquires a wound. However, both curves are rather different
behaviors: when the contact term is included the node appears very
close to zero and the correlation function has a quite flat
behavior. This is a sign that  the correlation function used was not
flexible enough to respond simultaneously to the repulsive contact
interaction and to the attractive component of the interaction.

We have computed the binding energies as function of the density for
the Hartree, HF and HF+Corr and compared with the
quark-meson-coupling model (QMC) \cite{qmc}, as can be seen from
fig. 2. The inclusion of correlations make the equation of state
(EOS) softer than Hartree or HF calculations, if the contact term of
the pion contribution is neglected. We also see that the inclusion
of the $\rho$ and $\pi$-mesons brings extra softness to the EOS.
However, the curve which describes the model with contact term and
with the correlations taken into account shows a very stiff
behavior. This is due to the fact that a very simplified
parametrization of the correlation function was used, which was not
flexible enough to respond to the repulsive and to  attractive
components of the interaction. A softer EOS around nuclear matter
saturation density is also provided by QMC.

In Fig. 3, we plot the effective mass versus density of  nuclear
matter. If correlations are included and the  contact term is
neglected the effective mass does not decrease so fast with the
increase of density as in a Hartree or, even worse, HF calculation.
This explains the softer behavior of the EOS with correlations.
However, the variation of the effective mass with density is still
smaller within the QMC model.

Correlations also affect the behavior of the neutron matter equation
of state (EoS), which is plotted in Fig.4. In this calculation we
include the four mesons $\sigma,\, \omega,\, \rho$ and $\pi$. We
perform the calculation with and without short range correlations
and  with and without the delta term of the $\pi$ contribution.  The
delta term  makes the equation of state very hard in both
calculations: with and without short range correlations. We have
already discussed that the correlation function should be more
flexible to deal with the delta term. Both EoS without the delta
term show an unrealistic behavior: either binding in the Hartree
Fock calculation or a shallow minimum in the HF plus correlations
calculation. Neutron star observation data are compatible with a
zero density surface which would not be the case if a minimum at
finite density would occur in the neutron star EoS. It is, however
important to point out that the inclusion of correlations almost
lifts this  behavior of the neutron EoS typical of the Walecka model
\cite{chin}.

We stress that a node occurs in the relative wave function, and the
EOS becomes softer, when the energy is optimized with respect to
variations of $\alpha$. However, if $\alpha$ is set equal to 1, so
that the node in the relative wave function is replaced by a wound,
the EOS remains stiff. A variational treatment is therefore
essential.

We conclude that the explicit introduction of correlations has
important effects. The behavior of the EOS and the values of the
effective coupling constants are most sensitive to the presence of
short range correlations, both when we keep and when we omit the
contact term in the pion interaction. Finally, we observe that the
presence of flexible short range correlations tends to soften the
EOS. In fig. 2, the curve ``HF-corr-with contact term'' appears to
contradict this statement. This is because the correlation function
used was not flexible enough to respond simultaneously to the
repulsive contact interaction and to the attractive component of the
interaction. It is also clear that a richer parametrization of the
correlation function, such as $f(r)=1-(1+\alpha r+\beta
r^2)~e^{-\gamma r}$, is required if the contact term is included.
Work in this direction are in progress.

It should be said that conclusions drawn from a study of this kind
have only qualitative strength since the healing distance constraint
imposed on the parameters of the correlation function is not
completely unambiguous and since higher order terms of the cluster
expansion of the expectation values have not been estimated.

\acknowledgments Valuable discussions with S. Moszkowski are
gratefully acknowledged. This work was partially supported by FCT
(Portugal) under the projects POCTI/FP/FNU/50326/2003, and
POCTI/FIS/451/94. PKP is grateful for the friendly atmosphere at
Department of Physics, University of Coimbra, where this work was
partially done.

\section{Appendix}
 The angular integrals are given by
\[A_i(k,k')=B_i(k,k')=2\pi ~\frac{g_i^2}{4\pi}\int_0^\pi d \cos\theta
 ~\tilde F_i(k,k',\cos \theta),\]
\[C_i(k,k')=2\pi~\frac{g_i^2}{4\pi}\int_0^\pi \cos \theta ~d \cos
\theta ~\tilde F_i(k,k',\cos \theta),\]
\[I(k,k')=2\pi \int_0^\pi d
\cos \theta~\tilde C_1(k,k',\cos \theta),\] and
\[J(k,k')=2\pi\int_0^\pi \cos \theta ~d \cos \theta
 ~\tilde C_1(k,k',\cos \theta),\]
 where
 \[
 \tilde F_i(\vec k,\vec k')=\int \left[f(r)V_i( r)f( r)\right]~
 e^{i(\vec k-\vec k') \cdot \vec r}~d\vec r \quad \quad
 \mbox{and}\quad\quad
 \tilde C_1(\vec k,\vec k')=\int (f2(r)-1)~
 e^{i(\vec k-\vec k') \cdot \vec r}~d\vec r.\]

 \begin{figure}[b]
 \includegraphics[width=.5\textwidth,angle=0]{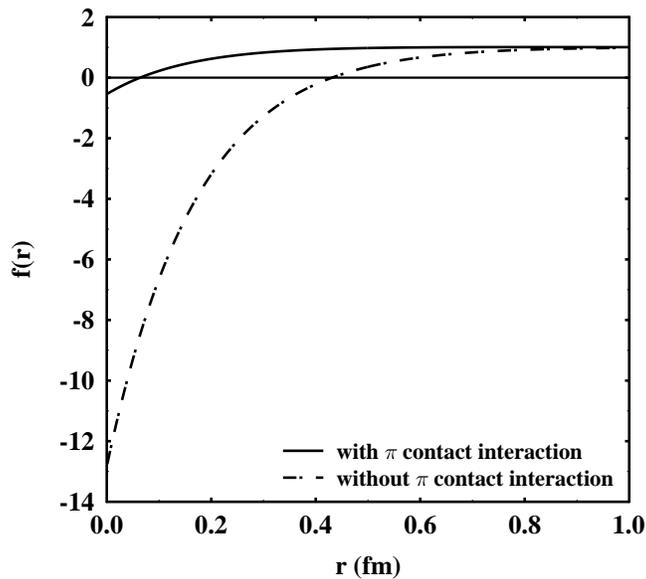}
 \caption{The correlation function $f(r)$ for the calculation with and without
  the delta term of the $\pi$ contribution.} \label{fig1}
 \end{figure}
 \begin{figure}
 \includegraphics [width=.5\textwidth,angle=0]{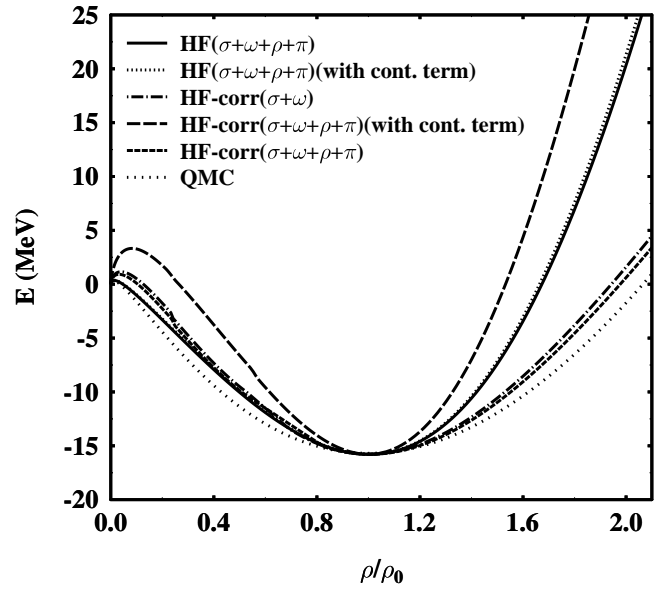}\caption{The
 equation of state with and without short range correlations.}
 \label{eos}
 \end{figure}
 \begin{figure}
 \includegraphics[width=.5\textwidth,angle=0]{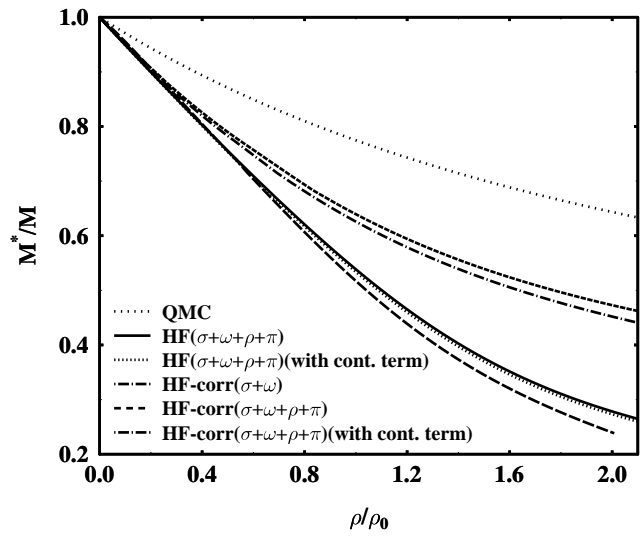}
 \caption{Effective mass as a function of density.} \label{fig4}
 \end{figure}
 \begin{figure}
 \includegraphics[width=.6\textwidth,angle=0]{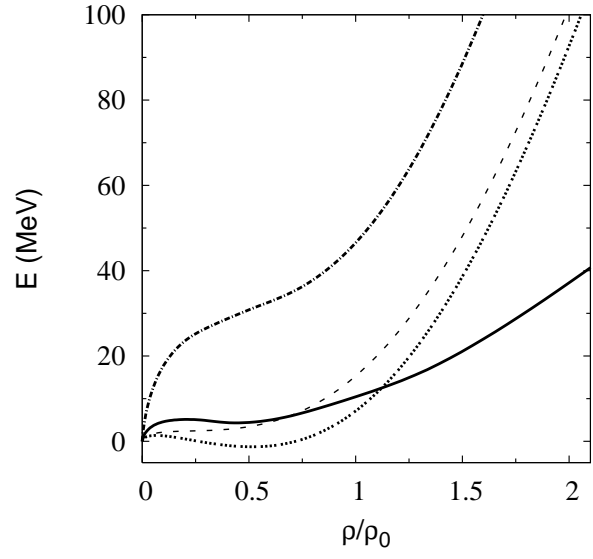}
 \caption{Equation of state of neutron matter (solid line: $\sigma+\omega+\rho+\pi+ \mbox{correlation (without delta)}$,
dash-dotted line:  $\sigma+\omega+\rho+\pi+ \mbox{correlation (with delta)}$,
dotted line:  $\sigma+\omega+\rho+\pi$ (without delta),
dashed line: $\sigma+\omega+\rho+\pi$ (with delta)} \label{fig5}
 \end{figure}

 \begin{table}
 \begin{ruledtabular}
 \caption{Parameters of nuclear matter. We have used a density
 dependent parameter (HF+corr) $\gamma=600+400~k_F/k_{F0}$ MeV for
 the correlation. These parameters were obtained with fixed: $M=939$
 MeV, $m_\sigma=550$ MeV, $m_\omega=783$ MeV, $m_\rho=770$ MeV,
 $m_\pi=138$ MeV, $g_\rho^2/4\pi=0.55$ and $f_\pi^2/4\pi=0.08$ at
 $k_{F0}=1.3$ fm$^{-1}$ with binding energy
 $E_B=\varepsilon/\rho-M=-15.75$ MeV. The value of $\gamma$ refers to
 saturation density.} \label{tab}
 \begin{tabular}{cccccc}
 &$g_\sigma$ & $g_\omega$&$\alpha$& $\beta$& $\gamma$\\
 & & & & (MeV) & (MeV)\\
 \hline Hartree &11.079&13.806&  &  &\\
 HF &10.432&12.223&  &  &\\
 HF+corr$(\sigma+\omega)$&4.4559&2.6098& 13.855 &-2252.448 &1000\\
 HF+corr$(\sigma+\omega+\rho+\pi)$&3.1925&2.199& 13.822 &-2258.037&1000\\
 HF+corr$(\sigma+\omega+\rho+\pi+\pi (\mbox{contact
 term}))$&11.662&12.930& 1.5435 &-497.391&1000
 \end{tabular}
 \end{ruledtabular}
 \end{table}
 \begin{table}
 \begin{ruledtabular}
 \caption{ Ground state properties of nuclear matter at saturation
 density.  } \label{tab1}
 \begin{tabular}{ccccccc}
 & $M^*/M$& ${\cal T}/\rho_B -M$& ${\cal V}_d/\rho_B$&
 ${\cal V}_e/\rho_B$&${\cal T}^C/\rho_B$& contact term \\
 & & (MeV)&(MeV)&(MeV)&(MeV)& (MeV)\\
 \hline Hartree & 0.540 & 8.11 &-23.86 &  & &\\
 HF & 0.515 & 5.87 &-37.45 &15.83& &\\
 HF+corr $(\sigma+\omega)$&0.625& 15.95&-73.12 &20.46&19.96&\\
 HF+corr $(\sigma+\omega+\rho+\pi)$&0.645& 16.41&-17.76 &-34.57&20.16&\\
 HF+corr $(\sigma+\omega+\rho+\pi+\pi (\mbox{contact term}))$&0.517&
 6.67&-42.30 &11.11&2.53 & 6.21
 \end{tabular}
 \end{ruledtabular}
 \end{table}
 \end{document}